\documentclass{article}

\usepackage{authblk}

\usepackage{graphicx}
\usepackage{multirow}
\usepackage[numbers,sort]{natbib}
\usepackage{caption}

\let\oldtabular\tabular
\renewcommand{\tabular}{\footnotesize\oldtabular}

\begin{document}

\title{An atomic hydrogen beam to test ASACUSA's apparatus for antihydrogen spectroscopy}


\author[1]{\mbox{M. Diermaier}}         
\author[1]{\mbox{P. Caradonna}	} 
\author[1]{\mbox{B. Kolbinger}}	 
\author[2]{\mbox{C. Malbrunot}}		
\author[1]{\mbox{O. Massiczek}}		 
\author[1]{\mbox{C. Sauerzopf}}		 
\author[1]{\mbox{M. C. Simon}}		
\author[1]{\mbox{M. Wolf}}	 
\author[1]{\mbox{J. Zmeskal}}
\author[1]{\mbox{E. Widmann}}
\affil[1]{Stefan Meyer Institute for Subatomic Physics, Austrian Academy of Sciences\\ 1090 Vienna, Austria}
\affil[2]{CERN, 1211 Geneva 23, Switzerland
}
\date{}


\maketitle

\begin{abstract}
The ASACUSA collaboration aims to measure the ground state hyperfine splitting (GS-HFS) of antihydrogen, the antimatter pendant to atomic hydrogen. 
Comparisons of the corresponding transitions in those two systems will provide sensitive tests of the CPT symmetry, the combination of the three discrete symmetries charge conjugation, parity, and time reversal.
For offline tests of the GS-HFS spectroscopy apparatus we constructed a source of cold polarised atomic hydrogen. 
In these proceedings we report the successful observation of the hyperfine structure transitions of atomic hydrogen with our apparatus in the earth's magnetic field.

\end{abstract}

\section{Introduction}
\label{intro}

The hydrogen (H) atom is one of the most precisely investigated systems in physics. For instance, the 1s-2s transition has been measured with a precision of
\mbox{$<10^{-15}$ \cite{Ref1}} and the GS-HFS is known to a level better than $10^{-12}$ \cite{Ref2}.

As a consequence high sensitivity to CPT violations can be achieved by measuring the same transitions in antihydrogen $\bar{\textrm{H}}$ \cite{Ref8}, since CPT invariance predicts that the properties of particles and their antiparticles are exactly the same or exactly the opposite.

The antihydrogen group within the ASACUSA collaboration focuses its experiments on testing the CPT symmetry by comparing the GS-HFS of the hydrogen and antihydrogen atom \cite{Ref3}.
During the Long Shutdown 1 period at CERN antiprotons could not be delivered and therefore no $\bar{\textrm{H}}$s could be produced. Hence we constructed a source of cold polarised atomic hydrogen to test and characterise components of the spectroscopy beam line, that will be used for the $\bar{\textrm{H}}$-HFS measurement. The main two components are a radio~frequency~(RF) cavity to induce spin flips and a superconducting sextupole magnet used as a spin state analyser.

\section{Method}
\label{sec:1}
\begin{figure*}

  \includegraphics[width=\textwidth]{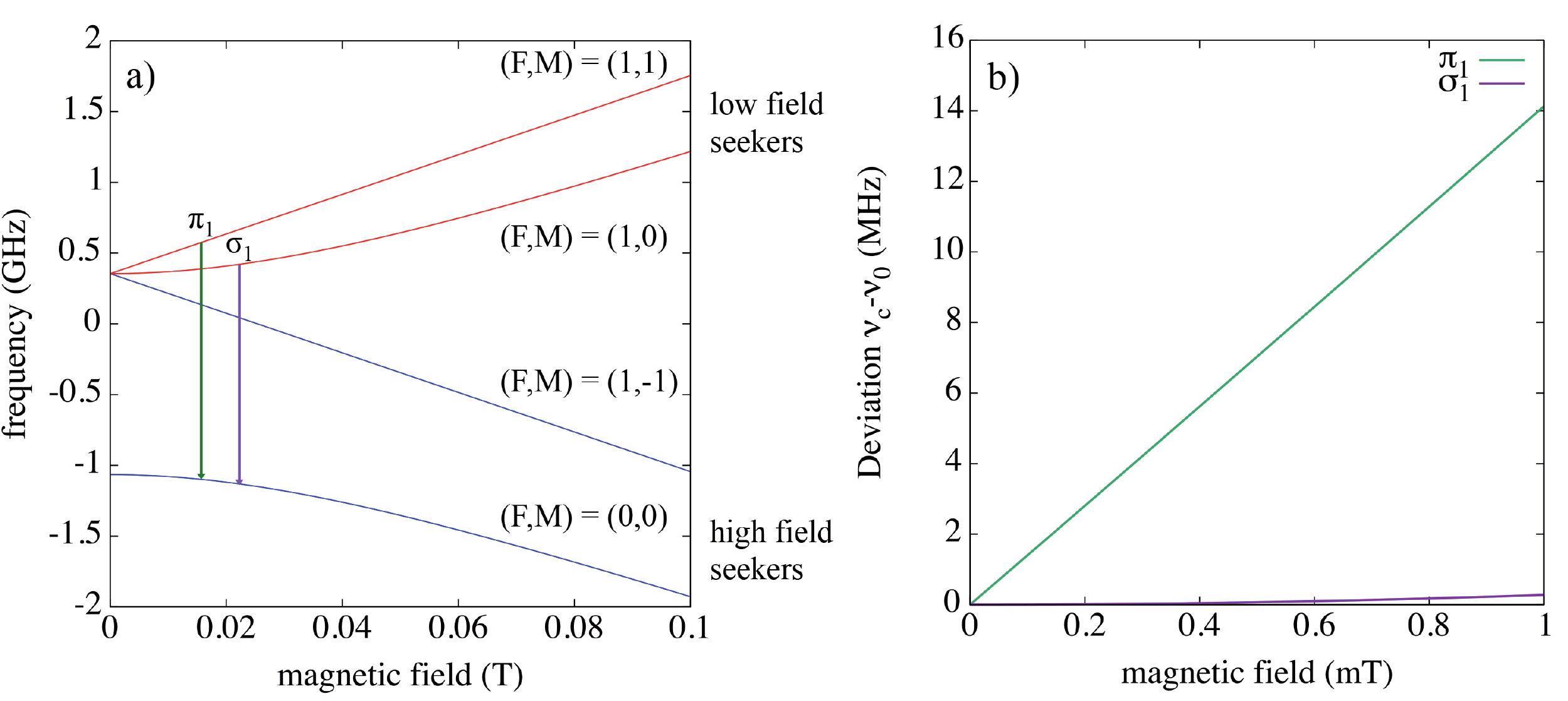}

\caption{a) Breit Rabi diagram for hydrogen: The interaction of the electron and the proton can be described with the quantum numbers $F$, the total spin, and $M$, the projection of $F$.
$(F,M) = (0,0)$ is a singlet state and $(F,M) = (1,-1), (1,0), (1,1)$ a triplet state.
The transition from $(F,M) = (1,1)$ to $(0,0)$ is called $\pi_1$ and the transition from $(F,M) = (1,0)$ to $(0,0)$ is called $\sigma_1$ \cite{Ref6}.
b) Deviation of $\pi_1$ and $\sigma_1$ transition frequencies ($\nu_c$) in magnetic field. The $\pi_1$ transition is shifting faster than the $\sigma_1$ transition and therefore this transition is more sensitive to magnetic field inhomogeneities.}
\label{fig:1}       
\end{figure*}

HFS transitions can be directly measured by the magnetic resonance technique invented by Rabi in the 1930s \cite{Ref7}. The best measurement of the hydrogen GS-HFS using Rabi's method achieved a precision of $4 \times10^{-8}$ \cite{Ref9,Ref10}. Higher precision has been reached in maser experiments \cite{Ref12,Ref13}, however, this method is not directly transferable to $\bar{\textrm{H}}$.
Rabi's method is based on spatial separation of spin states in external magnetic field gradients. As can be seen in figure~\ref{fig:1}, the triplet splits up into two states $(F,M) = (1,1), (1,0)$ that increase their energy in presence of an external magnetic field, while the third triplet state $(F,M) = (1,-1)$ lowers its energy like the singlet state $(F,M) = (0,0)$. As a result, in two states the atoms experience a force toward higher magnetic fields (high field seekers) while in the other two states the atoms experience a force toward lower magnetic field (low field seekers).

\begin{figure*}

  \includegraphics[width=\textwidth]{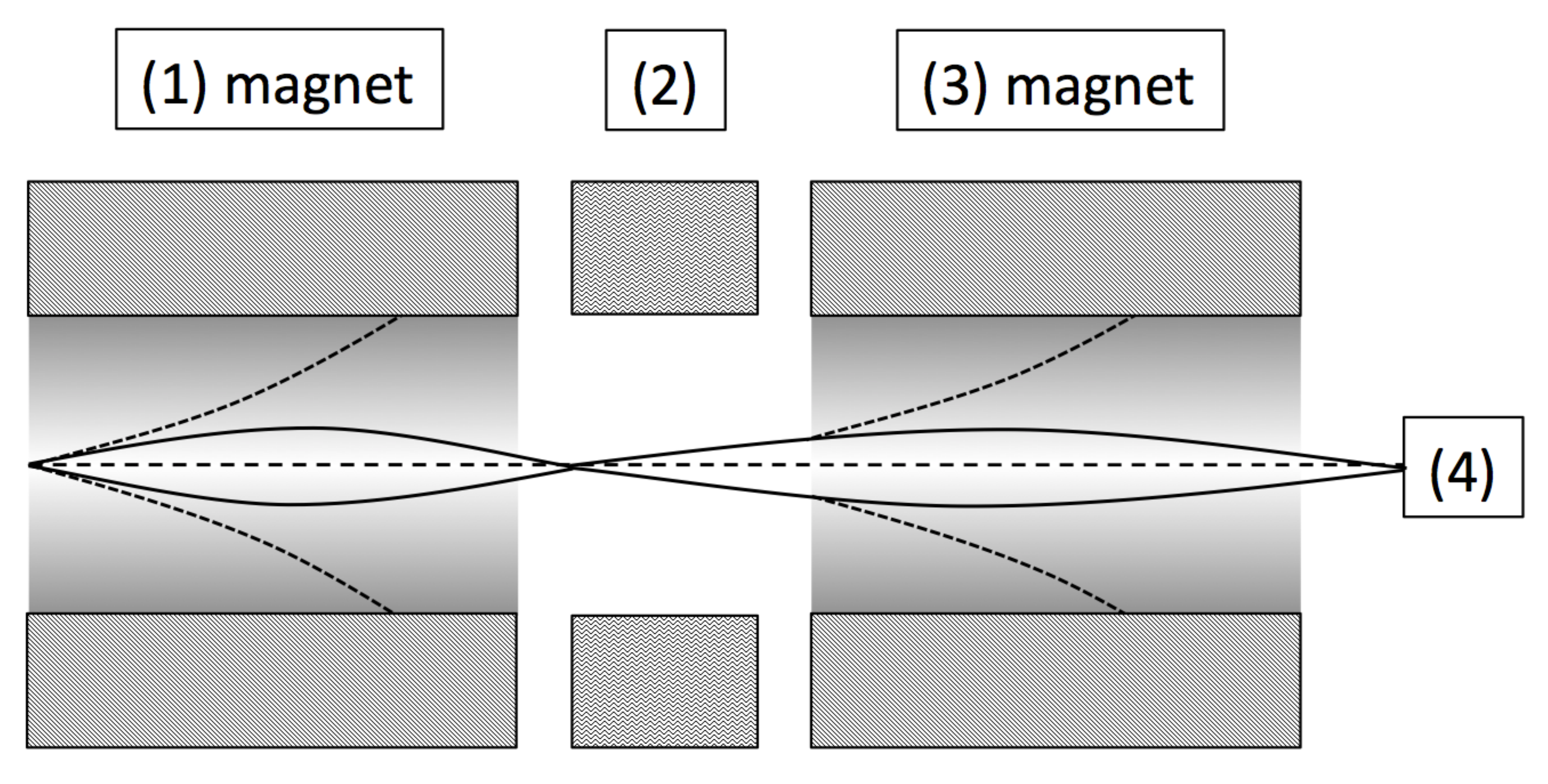}

\caption{Illustration of the Rabi experiment \cite{Ref7}: In the magnet (1) a beam of hydrogen atoms gets polarised via Stern-Gerlach separation. In the setup described in this paper, sextupole magnets are used for the separation. 
The solid lines represent the low field seekers, the short dashed lines represent the high field seekers which are deflected, and the long dashed lines show the beam centre.
In section (2) an oscillating microwave field with a frequency of about 1.42 GHz induces spin flips. In section (3) a second magnet  analyses the beam. If spin flips occur, low field seekers are converted into high field seekers and therefore a drop in counting rate should be seen at the detector (4).}
\label{fig:2}      
\end{figure*}

A Rabi type experiment employs the following main components as identified in figure \ref{fig:2}:

(1) a B field gradient to generate a spin polarised beam, (2) an oscillating B-field to drive spin flips, (3) a second B-field gradient to select not flipped spins, and (4) a detector. By scanning the frequency of the oscillating B-field the number of spin flips will vary leading to a drop in count rate on the detector when spin flips are occurring.

\section{Experimental setup}
\label{sec:2}

The aforementioned main components of a Rabi-type experiment are used in our $\bar{\textrm{H}}$ and H experiments. 
In the $\bar{\textrm{H}}$ experiment the polarised $\bar{\textrm{H}}$ beam will emerge from a cusp trap, where the mixing of $\bar p$ and $e^+$ takes place \cite{Ref3}.
Spin flips are induced in a strip-line cavity with a central frequency of 1.42 GHz and a Q-value of about 100 \cite{Ref14}. A superconducting sextupole magnet (Tesla Engineering), with a maximum pole field of 3.6 T selects the atoms in its field gradients, by focussing low field seekers and defocussing high field seekers. $\bar{\textrm{H}}$ events are counted afterwards in an annihilation detector downstream of the superconducting magnet.

\begin{figure}
  \includegraphics[width=\textwidth]{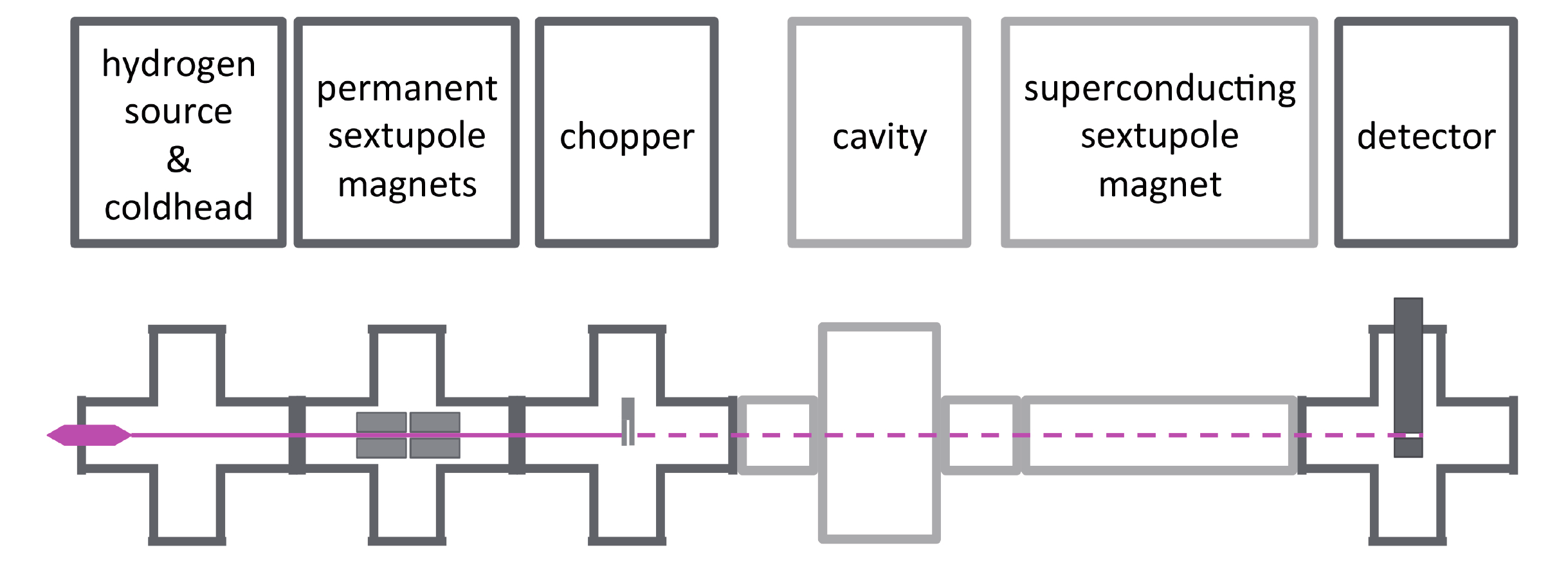}
\caption{H-Beam line: atomic hydrogen is formed in a plasma and then cooled with a coldhead. Sextupoles polarise the beam by sorting out the high field seekers. A chopper modulates the beam for the usage of background suppression with a lock-in amplifier. In the cavity spin flips can be driven. The superconducting sextupole magnet analyses the spin states and finally a quadrupole mass spectrometer detects the beam.}
\label{fig:3}   
\end{figure}

Atomic hydrogen, on the other hand, is produced by dissociation of molecular hydrogen in a microwave driven plasma contained in a pyrex glass cylinder (see ref. \cite{Ref4}). 
The atoms enter a vacuum chamber through a small orifice in the pyrex cylinder and a PTFE tubing which is kept at cryogenic temperatures in order to cool the beam to typically 50 K. The beam is polarised by using permanent sextupole magnets that have a inner diameter of 1 cm, a length of 6 cm, and a maximum pole field of 1.3 T.
The same components for spin flipping and subsequent spin state selection are used for both, the H and $\bar{\textrm{H}}$ experiment. To detect the H beam we use a MKS Microvision 2 quadrupole mass spectrometer (QMS). Figure \ref{fig:3} illustrates schematically the beamline.

\section{Results and Discussion}
\label{sec:3}

In the presence of a static magnetic field such as the earth magnetic field the frequency for a $\pi_1$ and $\sigma_1$ transition shift towards higher frequencies as can be seen in figure \ref{fig:1}. 
The $\pi_1$ transition is more sensitive to field changes and inhomogeneities than the $\sigma_1$ transition. 

\begin{figure*}
  \includegraphics[width=0.75\textwidth]{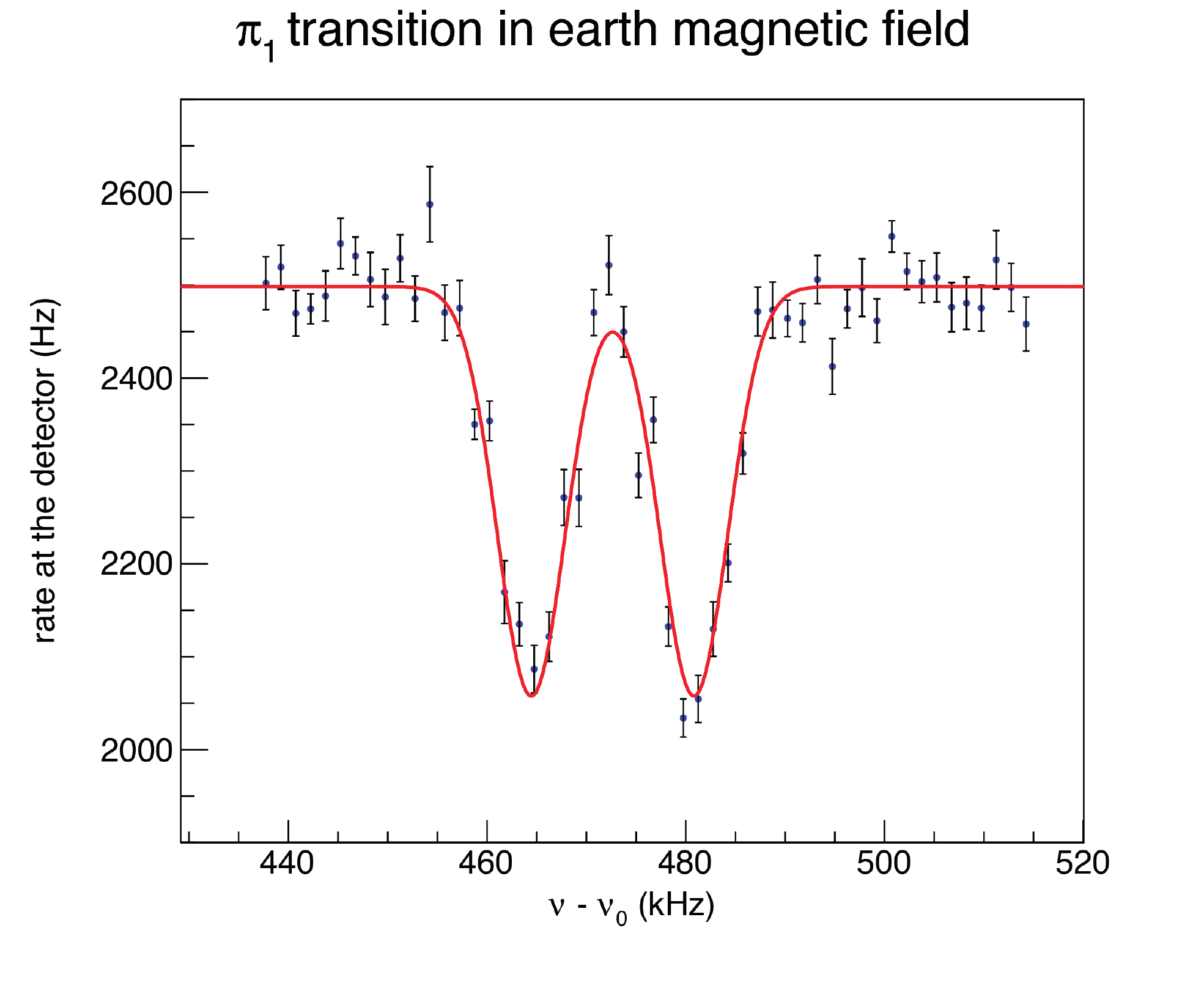}
  \includegraphics[width=0.75\textwidth]{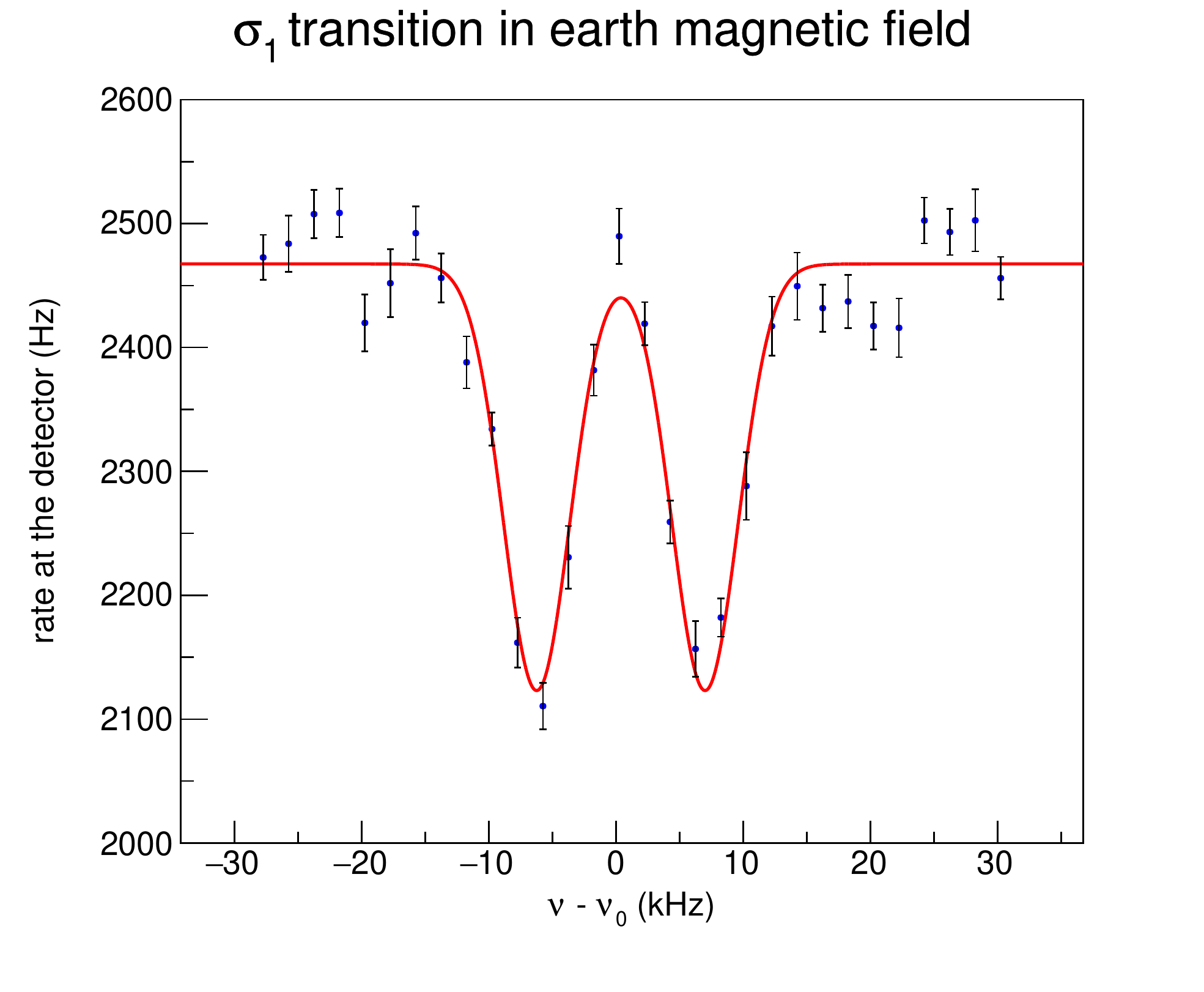}
\caption{$\pi_1$ and $\sigma_1$ transitions measured in the earth magnetic field with our spectrometer. The measured resonances were fitted with a double gaussian function. From the fit we get the standard deviation for  the single gaussian of the double gaussian function and the centre frequency.}
\label{fig:5}      
\end{figure*}

We were able to observe $\pi_1$ and $\sigma_1$ transitions without any magnetic shielding as can be seen in figure \ref{fig:5}.
The microwave cavity was driven with a Rohde and Schwarz SML02 synthesiser referenced to a Stanford Research FS725 Rb oscillator. The microwave signals were amplified by a Mini-Circuits ZHL-10W-2G(+) amplifier and her typically 1.3 W power was needed to drive a spin flip. The frequency was scanned by changing the synthesiser frequency and keeping the power constant.

The resonances in our cavity have a double peak structure. 
This line shape of the resonances is a consequence of the oscillating magnetic field inside the cavity that drives the spin flips. 
The standing waves have a sine distribution in beam direction and therefore at centre frequency we observe a maximum in counting rate, whereas at the side lobes we drive the spin flips.

The measurements for each resonance scan took a few hours. 
We obtained the signals with a lock-in amplifier that used the modulated signal of the chopper  and the signal from the mass spectrometer. The rate we observed is therefore the detected rate of the beam.
We fitted the data with two gaussians which gave a value for the centre frequency $\nu_c$ (see table \ref{tab:1} for the fit results). For the $\pi_1$ transition the deviation $\nu_c - \nu_0 = 472636 \pm 102$ kHz, for the $\sigma_1$ transition $\nu_c - \nu_0 = 379 \pm 95$ Hz. The errors are purely statistical and are coming from the fit. 
From the fit we also get the standard deviation for a single gaussian of the double gaussian function. This quantity is approximately a third bigger for the $\pi_1$ transition. This is a consequence of the sensitivity for $\pi_1$ which broadens the resonance. 
The magnetic field was measured with two Bartington Mag-03IE1000 flux gates mounted on
the outside of the cavity body on the upstream and downstream side.
Due to magnetic field inhomogeneities the two sensors deviated from each
so that we can only estimate the magnetic field inside the cavity from the average and deviation of two measured values which gives a magnetic field of $37\pm4.2$ $\mu$T.
Using the Breit-Rabi formula one can also calculate, with the values from the centre frequency, the magnetic field that was present during the measurement. The values are also given in table \ref{tab:1} and are in agreement with the measured ones. 

\begin{table}
\caption{Results for $\pi_1$ and $\sigma_1$ transitions: First column: transition name. Second column: the centre frequency minus literature value. The errors are statistical and are coming from the fit. Third column: standard deviation (stdev.) of a single gaussian from the fit function. Fourth column: magnetic field calculated with Breit-Rabi formula using centre frequency of the fit. The error is gained from error propagation.
Fifth column: the measured magnetic field.}
\label{tab:1}       

\begin{tabular}{p{1.25cm} p{1.75cm} p{1.95cm} p{2.7cm} p{2.2cm} lllll}
\hline\noalign{\smallskip}
transition & $ \nu_c-\nu_{0}$ (Hz)  & stdev. of single gaussian (Hz) & calculated $\newline$ magn. field ($\mu$T) & measured $\newline$ magn. field ($\mu$T)\\
\noalign{\smallskip}\hline\noalign{\smallskip}
$\sigma_1$ & 379 $\pm$ 102 & 2606 $\pm$ 107 &  37 $\pm$ 5 &\multirow{2}{*}{$37 \pm 4.2$} \\
$\pi_1$ & 472636 $\pm$ 95 & 3405 $\pm$ 92 & 33.770 $\pm$ $7 \times 10^{-3}$ \\
\noalign{\smallskip}\hline
\end{tabular}
\end{table}

\section{Conclusion and  Outlook}
\label{sec:4}
In this work we presented the first successful observation of both $\sigma_1$ and $\pi_1$ transitions obtained in our setup in the earth's magnetic field, without using any magnetic shielding. The measurements are a clear indication that Majorana spin flips, described in \cite{Ref15} are not a serious problem for our method.
The values of the measured HFS frequencies agree within error to those calculated from the Breit-Rabi formula using the measured magnetic field.
The observed line width of $4\times10^{-6}$ and the statistical precision of the line centre of $~10^{-7}$ promises to enable us to determine the hydrogen HFS with significantly higher precision by applying well defined and characterised external magnetic fields and performing systematic studies of their effects.

\subsubsection*{Acknowledgements}

The authors would like to thank H. Knudsen and Hans-Peter Engelund Kristiansen for providing the hydrogen source, C. Klaushofer for his helping hands, and the CERN cryolab for providing their laboratory. This work is supported by the European Research Council grant no. 291242-HBAR-HFS and the Austrian Ministry for Science and Research.

\end{document}